\documentclass{article}
\usepackage[utf8]{inputenc}
\usepackage{amsmath}
\usepackage{amssymb}
\usepackage{amsfonts}
\usepackage{amstext}
\usepackage{amsthm}
\usepackage{authblk}
\usepackage{xcolor} 

\usepackage{tikz,lipsum,lmodern}
\usepackage[most]{tcolorbox}

\usepackage[normalem]{ulem} 

\title{Mesoscale Modeling of an Active Colloid’s Motion}
\author[$\dagger,*$]{Matthew Dobson}
\author[$*$]{David Masse}
\date{\today}
\affil[$*$]{\small{Department of Mathematics and Statistics, University of Massachusetts Amherst, USA}}
\affil[$\dagger$]{\small{Corresponding author: dobson@umass.edu}}

\begin{document}

\maketitle

\begin{abstract}
This paper uses Cahn-Hilliard equations as a mesoscale model of the motion of active colloids.  The model attempts to capture the driving mechanisms and qualitative behavior of the isotropic colloids originally proposed by J. Decayeaux in 2021. We compare our model against the single colloid behavior presented in that work, as well as against multi-colloid systems. 
\end{abstract}

\section{Introduction}
Active colloids have received continued research interest due to their potential utility in a variety of fields including biomedical~\cite{Kaew1}, materials  science~\cite{Duan1}, and manufacturing, particularly for polymers~\cite{Walt1}. Their dynamics have been explored both numerically and through the derivation of equations of motion from first principles~\cite{Ram1,Lan1,Lieb1,Robert1}. The primary characteristic of active colloids is their enhanced diffusion compared to inactive ones. A variety of mechanisms exist to produce this effective propulsion, such as the miniature motors used by micro-organisms and the chemically induced propulsion of Janus spheres. These mechanisms continue to be studied both in experiment and simulation~\cite{Bish1,Ebens1,Su1}. A common theme among these mechanisms is that they require some asymmetry in the colloid or propulsion mechanism to produce the enhanced diffusion effect. 

Recently, a microscopic model of an isotropic colloid has also been demonstrated to achieve enhanced diffusion. This work by Decayeux et al.~\cite{Deca21,Deca2} demonstrates enhanced diffusion with numerical simulations and posits effective equations of motion. Despite being isotropic, the Brownian motion of the colloid induces a local phase separation in the surrounding solute, which promotes directed motion and leads to the enhanced diffusion. In the original work, only a single colloid was simulated. Previous research on other active colloids has focused on clustering behavior when large numbers of colloids are present, as in ~\cite{Bialke1}. This behavior has been modeled using the Cahn-Hilliard equations for example in ~\cite{Speck1}. In the present work, we present a model for the dynamics of low density isotropic colloids using the Cahn-Hilliard equations. We will demonstrate that our model captures the enhanced diffusion of a single colloid as well as compare against multi-colloid systems. 

\section{Problem Description}
The model used by Decayeux et al. consists of a number of identical solute particles and a single colloid particle immersed in a bath in a square two dimensional domain with periodic boundary conditions. All particles are subject to overdamped Langevin dynamics with interactions mediated by either a Weeks-Chandler-Anderson (WCA) or Lennard-Jones (LJ) potential. Let $\mathbf{r}_i$ be the position of particle $i$, 
$U(\mathbf{r}_i - \mathbf{r}_j)$ be the potential between two particles, and $\mathbf{\eta}_i$ the Brownian White noise acting on particle $i$, then the overdamped Langevin dynamics of the system are written as~\cite{Deca21}:
\begin{equation}
\mathbf{\dot{r}}_i(t) = -\frac{D_i}{k_B T}\sum_{i\neq j}\nabla U(\mathbf{r}_i - \mathbf{r}_j) +\sqrt{2D_i}\mathbf{\eta}_i (t)
\end{equation}
Using a forward Euler integration scheme, the update rule is given by:
\begin{equation}
\mathbf{r}_i(t + \Delta t) = \mathbf{r}_i(t) - \frac{D_i}{k_B T}\sum_{i\neq j}\nabla U(\mathbf{r}_i - \mathbf{r}_j)\Delta t +\sqrt{2D_i\Delta t}\mathbf{\eta}_i
\end{equation}
Here $D_i$ is the bare diffusion coefficient of the particle, representing its diffusivity if it were the only particle in the bath. The parameters $k_B$ and $T$ are the Boltzmann Constant and temperature respectively. The diffusion coefficient is: 
\begin{align*}
D_i = &  \begin{cases} 
    1/5 & \text{Colloid ($D_S$)} \\
    1 & \text{Solute, ($D_C)$}
    \end{cases}
\end{align*}

The pairwise inter-particle potential depends on the types of particles involved. Initially all solute particles are of type `A' and repulse each other. However, the colloid is endowed with a circular region of influence around it that transforms, at a chosen rate, the solute particles into type 'B' particles which attract one another. Interactions between differing solute particle types are still repulsive, and both types repulse the colloid. Our present simulations will have multiple colloids, and these also repulse each other in the same manner. In all cases the repulsive forces arise from the WCA potential. Outside the region of influence, particles return to being type 'A' again at a chosen rate. The reaction rates are assumed to be identical, and are taken to be 10 per unit time, with reaction radius of 7.5 length units. The attractive force between two type 'B' solute particles is mediated by the LJ potential. We may write these potentials in the following forms:
\begin{equation}
U_{WCA}(r_{ij}) = 4\epsilon' \left[\left(\frac{d_{ij}}{r_{ij}}\right)^{12} - \left(\frac{d_{ij}}{r_{ij}}\right)^6 \right] + \epsilon'
\end{equation}
\begin{equation}
U_{LJ}(r_{ij}) = 4\epsilon \left[\left(\frac{d_{ij}}{r_{ij}}\right)^{12} - \left(\frac{d_{ij}}{r_{ij}}\right)^6 \right]
\end{equation}
We set $\epsilon'=10k_B T$ and $\epsilon=3k_B T$. The same values are used for all interactions. Finally $d_{ij} = (\sigma_i + \sigma_j)/2$, where $\sigma$ represents the particle diameter. In our simulations the solute particle diameter, also taken to be the characteristic length scale of the system, is 1, and for colloids it is 5. 

The region the simulation is taking place in is a square of 100 length units in side length, with periodic boundary conditions. The characteristic time of the model is taken to be $\tau=\sigma_S^2/D_S$, with time steps $\Delta t=0.00003\tau$, and total time being $1500\tau.$ Due to the physical size of the colloids taking up considerable area, in our simulations we seek to keep the solute density constant in the region not occupied by colloids. For simulations with a single colloid, we use 1000 solute particles. The most colloids simulated is 128, and in this case there are 750 solute particles in the simulated area.

The original study demonstrated that the colloid in this setup could experience enhanced diffusion compared to a non-reactive colloid so long as the above parameters were within a certain range. Furthermore, the effective equations of motion of the colloid matched those of a standard active Brownian particle~\cite{Deca21}.  The model has a number of parameters that one might want to experiment with.  For example, if the density of solute particles in the domain is high, then the colloid quickly becomes surrounded by a ring of solute particles and the enhanced diffusion effect is lost. In fact, the diffusion can be less than that of an inactive particle in an otherwise identical setup. Similarly if the reactive region of the colloid is very large, then solute particles will clump around each other but not necessarily in the region of the colloid, so their influence will be greatly diminished. See Appendix A for a comparison of a colloid with infinite reaction radius to one that is inactive. Another parameter not considered in the original paper is the number of colloids. While one colloid can be effectively modeled as an active Brownian particle, it is not the case that a collection of them can be modeled as independent Brownian particles. A natural question then is if there is a mesoscale model that can effectively capture the dynamics of this system over a wide range of parameters including number of solute particles. Here we propose a model using the Cahn-Hilliard equations, which are well suited for capturing the separation of the solute particles into high and low density regions. We will demonstrate that this model captures the behavior of single and multi-colloid systems, up to a limit in colloid density.

\section{Proposed Model}
The reactive region of the colloid has the effect of concentrating solute particles locally around it. Outside of this region the solute particles separate again and dissipate due to random motion. So the dynamics of our mesoscale model must be able to capture these concentrating and dissipating effects, and do so in relation to the changing position of the colloid particles. This focus on concentration inspires the choice of the Cahn-Hilliard equations to be a base for our model. The Cahn-Hilliard equations govern the phase separation of a binary mixture, and are typically written in terms of the concentration of one of the components. For our purposes we interpret this concentration as that of the solute particles. Recent work has investigated interactions between active colloids and a system governed by Cahn-Hilliard dynamics;~\cite{Diaz1}~\cite{Roy1} the novelty of our model is how we use the Cahn-Hilliard dynamics to drive the colloids themselves. Specifically, we couple the colloid's motion to the concentration gradient that it itself produces. In this way we have a model that is both local for an individual collloid and has natural interaction effects for multiple colloid systems.

\begin{figure}
\centerline{\includegraphics[scale=0.49,angle=0]{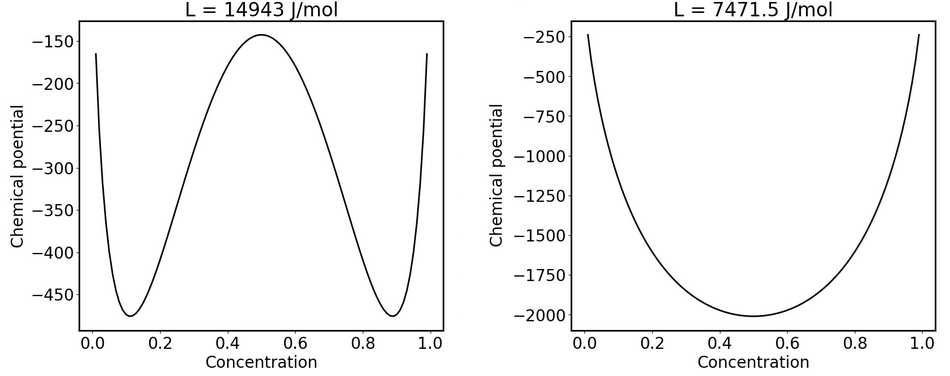}}
\caption{\label{fig:pot}\small Chemical potentials showing different stable regions. On the left, two stable regions exist, which drives the system to separation into high and low concentration regions. On the right, only one stable region exists, driving the system to a mixed state. The value of parameter L on the right is half of that for the figure on the left, so the dynamics of the system can be controlled by changing only this parameter.}
\end{figure}

We will use a form of these governing equations from ~\cite{Hill1,Yama1}, which will allow us to change the behavior close to colloid in a particularly simple way. Let $c$ be the concentration of solute particles, $\mu$ be the diffusion potential, and $M_c$ be the diffusive mobility, then we have the following differential equation governing the time evolution of the concentration:
\begin{equation}
\frac{\partial c}{\partial t} = \nabla\cdot\left(M_c(c)\nabla\mu(c,x)\right)
\end{equation}
Let us first focus on the diffusion potential $\mu$, which will depend on the distance to the colloid to induce the desired behavior in our model. Here $\mu$ is given by:
\begin{equation}
\mu(c,x)=RT\left[ \log(c)+\log(1-c)\right]+L(x)(1-2c)-a_c\nabla^2c
\end{equation}
Here $R=8.314\hspace{0.1cm}J/(mol K)$ is the Gas Constant, $T=673$K is the temperature. The parameter $a_c$ is called the gradient energy coefficient, and in our simulations $a_c = 3\times 10^{-14}\hspace{0.1cm}Jm^2/mol$. 
$L$ is a parameter that depends on the constituent particles, referred to as the atomic interaction parameter or regular solution constant ~\cite{Diaz1, Yama1}.
Typically this would be constant, but for our purposes $L=L(x)$ will depend on the position of the colloid, and will be the source of the reactive effect in our mesoscale model. To see why a varying $L$ gives us a simple description of the effect we are trying to reproduce, let us consider the chemical potential of our binary mixture:
\begin{equation}
g_{chem}(c)=RT\left[c \log(c)+(1-c) \log(1-c)\right]+L(x)c(1-c)
\end{equation}
Note that the first three terms of $\mu$ in equation (6) above are the partial derivative of $g_{chem}$ with respect to $c$ from equation (7). Let us define $L_0=14943\hspace{0.1cm}J/mol$ so that $L=L_0$ inside the colloid's region of influence, and $L=L_0/2$ outside of it. Calculating $g_{chem}$ with these values of $L$, we get the plots in Figure~\ref{fig:pot}, with $L=L_0$ in the left subplot and $L=L_0/2$ in the right subplot. The important feature here is that the left subplot has two stable regions that concentration can fall into, where as the right subplot has only one. By only varying $L$ we have gone from a system with two stable values of concentration to a system with only one.

 This gives us a natural tool to drive the concentration dynamics we want. In a circular region of radius 5.5 around the colloids, we simulate the Cahn-Hilliard equations with the $L=0.9L_0$ to promote phase separation into localized regions of high and low concentration. Outside this region at a distance greater than 7.5 units from a colloid, we use $L=L_0/2$ which drives the system back to a uniform concentration. We vary $L$ linearly from one value to the other near the intersection of these regions, the region between a radius of 5.5 and 7.5 from the center of the colloid. The choice of $L_0$ controls not only the final MSD but also the speed of phase separation, which is important for matching the particle models at intermediate times of ~100 time units. Finally we specify the form of $M_c$ as in ~\cite{Yama1, Zhu1}, which is a function of $c$ and diffusivity of the constituent particles. Our model has only one solute type, with diffusivity $D_S$, but since the colloids will be occupying the low concentration region it is a natural choice to interpret them as the second solute type and choose the other diffusivity to be $D_C$. Then $M_c$ is given by:
\begin{equation}
M_c(c)=\frac{D_S}{RT}\left[c+\frac{D_C}{D_S}(1-c)\right]c(1-c)
\end{equation}
Since our characteristic time scale is inversely proportional to $D_S$ and our dynamics are directly proportional to it, it's exact physical value doesn't play a role in the simulations.

The next step is to specify the interaction of the colloid with the concentration field. Before phase separation has occurred, the interaction is dominated by an effective damping of the Brownian motion that the colloid exposed to, due to the physical proximity of the colloid to the surrounding solute particles. This effect is captured in our model with a simple scale factor on the Brownian motion term in (2), with magnitude $\gamma = 0.88$ to match the MSD at small time scales. After phase separation has occurred, there is also an effective force in the direction of low concentration. This is captured with a simplified gradient following scheme. The gradient is calculated by summing up the change in concentrations around the perimeter of the colloid. The effective force the colloid receives is then this vector, which we will denote by $G_\Sigma$,  multiplied by a scale factor of $\beta = 0.021$ to match the MSD seen in the particle model. This is added to the force term in (2), giving in total:

\begin{equation}
\mathbf{\dot{r}}_i(t) = -\frac{D_i}{k_B T}\sum_{i\neq j}\nabla U(\mathbf{r}_i - \mathbf{r}_j) +\beta\,G_\Sigma+\gamma\sqrt{2D_i}\mathbf{\eta}_i (t)
\end{equation}

Equations (5-9) are our mesoscale model of the solute particles and the concentration dynamics, it remains to implement the colloid-colloid interactions. We use a soft potential that has the same effective distance as the WCA potential but which allows us to take larger time steps, as it doesn't diverge quickly when colloids overlap as the WCA potential does. The effective repulsive force on colloids is the same up to a distance of $1.06d_{Coll}$. After this point a polynomial force model is used that matches the magnitude and derivative of the force at $1.06d_{Coll}$ and brings the force back to zero at 0 distance. In practice both force models simply keep the colloids at the appropriate separation distance. The complete force model is illustrated in Figure~\ref{fig:Force1} and is given by is given by:

\begin{figure}
\centerline{\includegraphics[scale=0.5,angle=0]{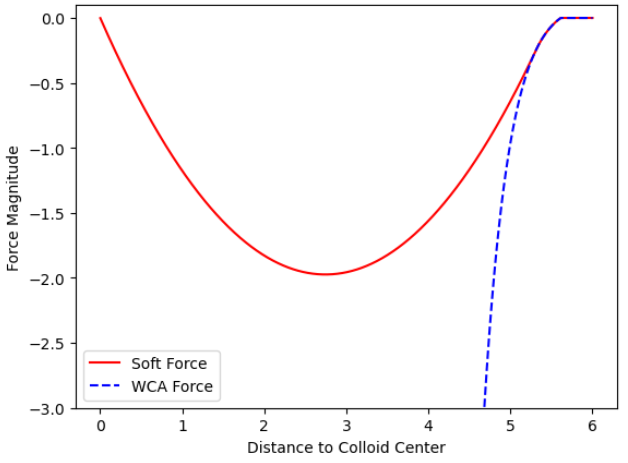}}
\caption{\label{fig:Force1}\small Illustration of Force model used in the mesoscale simulation compared to ordinary WCA.}
\end{figure}

\begin{align*}
\mathbf{F}_i = \sum_{i\neq j}  &  \begin{cases} 
    -\frac{D_C}{k_B T}\nabla U_{WCA}(\mathbf{r}_i - \mathbf{r}_j) & 1.06\,d_{C}\leq\|\mathbf{r}_i - \mathbf{r}_j\|\leq 2^{1/6}d_{C} \\
    -0.262\|\mathbf{r}_i - \mathbf{r}_j\|(\|\mathbf{r}_i - \mathbf{r}_j\|-5.4884) & 0\leq \|\mathbf{r}_i - \mathbf{r}_j\|< 1.06\,d_{C}
                               \end{cases}
\end{align*}

A final note about the model is that while the colloids have a physical size for the purposes of collisions, the concentration field still exists in the region that the colloid occupies, and the Cahn-Hilliard dynamics behave as normal there. Thus the complete model is represented by the following equations:
\begin{gather*} 
c(x,y,t+\Delta t) = c(x,y,t) + \left(\nabla\cdot\left(M_c(c)\nabla\mu(c,x)\right)\right)\Delta t \\
\mu(c,x)=RT\left[ \log(c)+\log(1-c)\right]+L(x)(1-2c)-a_c\nabla^2c\\
M_c(c)=\frac{D_S}{RT}\left[c+\frac{D_C}{D_S}(1-c)\right]c(1-c) \\
\mathbf{r}_i(t + \Delta t) = \mathbf{r}_i(t) + (\beta\,G_\Sigma +\mathbf{F}_i)\Delta t+\gamma\sqrt{2D_i\Delta t}\mathbf{\eta}_i
\end{gather*}

The simulation region is the same as in the particle model, and colloids have the same physical size of 5 units. The initial concentration field is 0.5325 with random fluctuations on the order of 0.01. The total simulation time is again $1500\tau$, with time steps of $0.1\tau$. Recall that time steps for the particle model were $0.00003\tau$, so the mesoscale model is certainly far more performant, a single trial taking on the order a minute to run compared to several hours for the particle model.

\section{Single Colloid Behavior}
\begin{figure}[ht]
\centerline{\includegraphics[scale=0.5,angle=0]{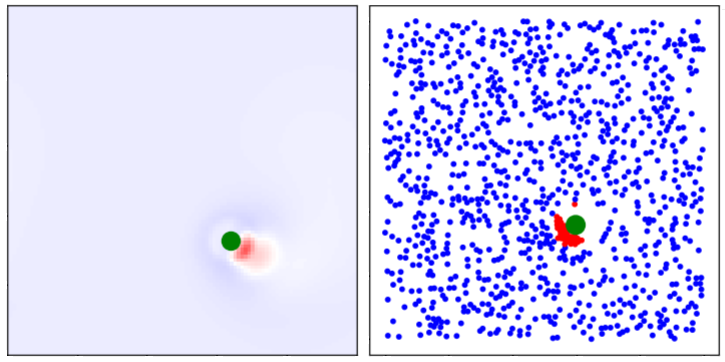}}
\caption{\label{fig:model1}\small High concentration regions effectively block the colloid from moving in that direction, so the otherwise random colloid motion becomes directed in the opposite direction. The left plot shows the particle setup, and the right shows the Cahn-Hilliard setup, both demonstrate the high concentration region that is the main driver of the enhanced diffusion.}
\end{figure}
Let us examine some simulation results that illustrate how our model captures the correct qualitative behavior of the system, particularly the mechanism of increased diffusion. In the right image of Figure~\ref{fig:model1}, we see a configuration of particles with the colloid in green, attractive solute particles in red, and repulsive solute particles in blue. Note the high density of red particles on one side of the colloid. As the colloid undergoes random motion, it will be obstructed from moving into this region by the collective repulsion of the mass of red particles. This directs the mean motion of the colloid in the opposite direction, upwards and to the right in this case. Similarly in the right image of Figure~\ref{fig:model1}, the colloid also undergoes random motion but is discouraged from moving into the high concentration red region, so its mean motion is directed upwards and to the left.

\begin{figure}
\centerline{\includegraphics[scale=0.42,angle=0]{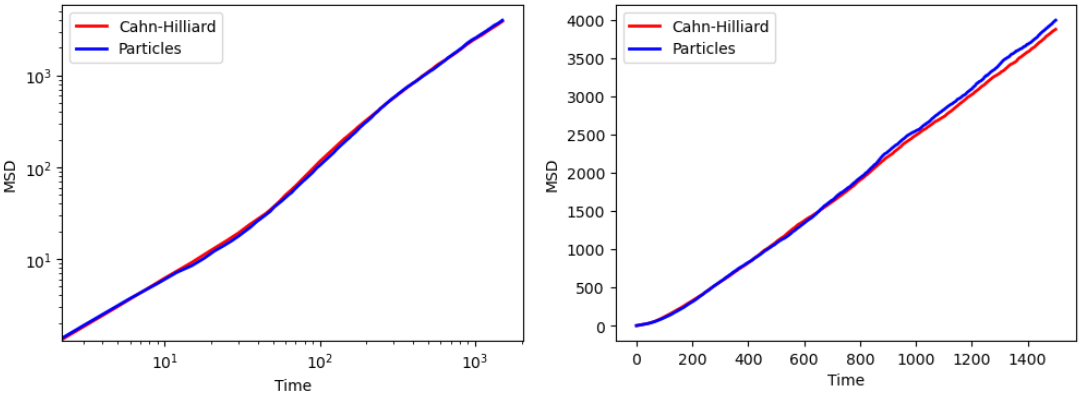}}
\caption{\label{fig:MSD1}\small Single colloid MSD comparison. The left plot is log-log scale and the right is linear scale. There is good qualitative and quantitative agreement across time scales.}
\end{figure}

A quantitative indicator of increased diffusivity of the colloid is the mean squared displacement (MSD). In Figure~\ref{fig:MSD1} we compare colloid MSD of the particle and mesoscale models. MSD is computed from the average of 1000 trials for a single colloid case and 500 trials multi-colloid cases for the particle model and an average of 2000 trials for the Cahn-Hilliard model. The log-log scale plot shows the characteristic shape of an active Brownian particle, with good qualitative and quantitative agreement. Though the particle model has slightly higher average MSD by the end of the simulation time, we expect this would be smoothed out closer to the Cahn-Hilliard model with a larger number of trials.
 
\section{Multiple Colloid Behavior}
So far we have seen that our model qualitatively captures the mesoscale behavior of a single colloid. Both the mesoscale model and particle model have no restriction to the number of colloids present, so a natural extension is to consider multiple colloids. To do so we first must specify what happens when regions of influence of the colloids overlap. In the particle case no extra considerations are needed, as we are only concerned if solute particles are inside or outside the region of influence. For the mesoscale model, the effect on the concentration field is governed by the closest colloid. 

\begin{figure}
\centerline{\includegraphics[scale=0.42,angle=0]{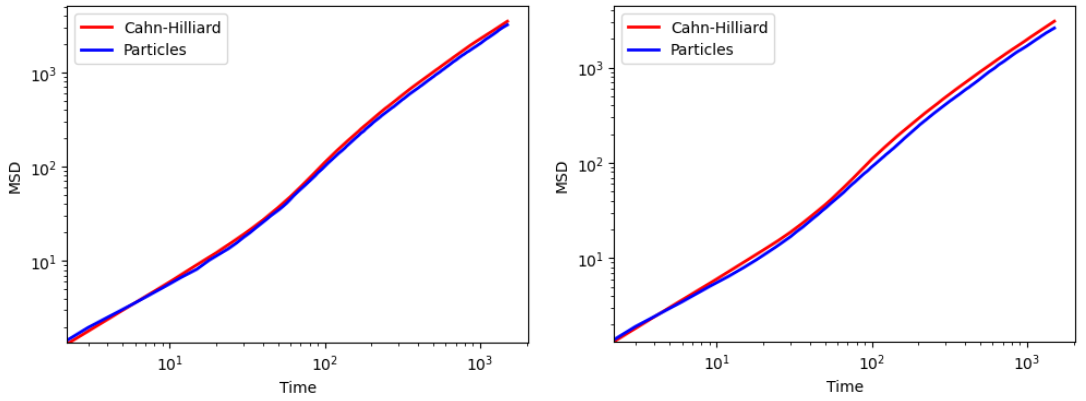}}
\caption{\label{fig:MSD4_8}\small MSD in log-log scale for four collolids (left) and eight colloids (right.)}
\end{figure}

\begin{figure}
\centerline{\includegraphics[scale=0.5,angle=0]{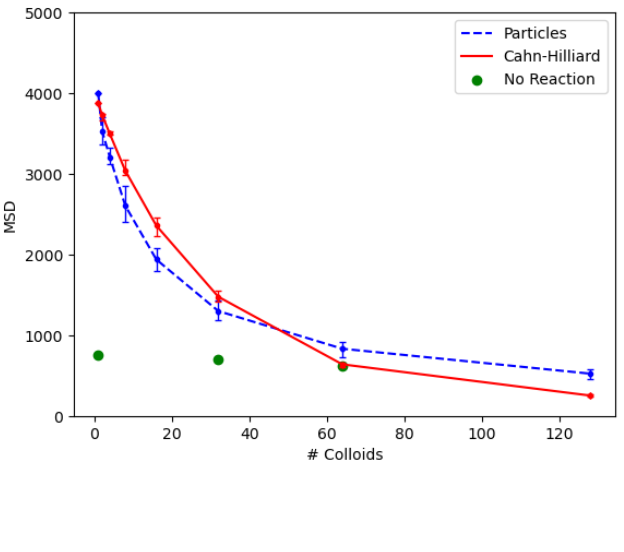}}
\caption{\label{fig:MSD_All}\small MSD at the end of the simulation time for $2^n$ colloids for in the range {0,7}. Added data points for systems with non-reactive colloids for comparison.} Error bars show range of MSDs averaged per colloid.
\end{figure}

We now compare MSD for simulations with a range of colloids in the simulation area. In the left subplot of Figure~\ref{fig:MSD4_8}, we see the results for four colloids. Because colloid positions are assigned randomly for every trial, we average together the results of all colloids for the multi-colloid trials. Total MSD is higher for the Cahn-Hilliard model by about 200, but qualitative agreement is still very good. In the right subplot we see results for eight colloids. Here the quantitative deviation in the Cahn-Hilliard model becomes apparent. The reason is that in the particle model colloids tend to cluster in larger numbers for slightly longer periods of time, leading to overall reduced MSD. Since all simulations are run for the same duration, we can compare MSD at the final time across the range of colloid quantities. This is shown in Figure~\ref{fig:MSD_All}. Additionally data points are shown for particle systems with the same physical setup but no reaction, showing that the enhanced diffusion effect exists even for denser systems. We see that although the Cahn-Hilliard model captures the overall downward trend of MSD, it does not capture the exact trend, with a change in behavior for dense systems.

\begin{figure}
\centerline{\includegraphics[scale=0.5,angle=0]{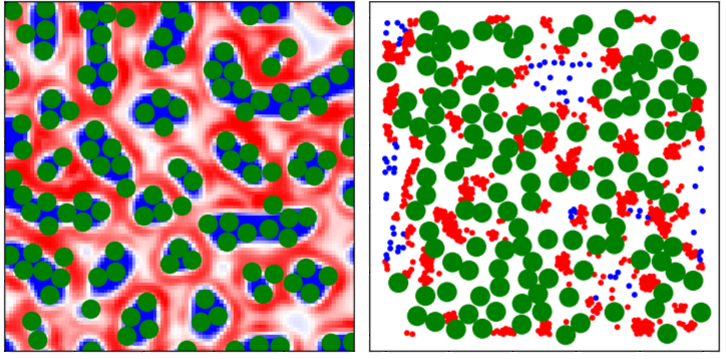}}
\caption{\label{fig:many_coll}\small Comparison of systems with 128 colloids. On the left is a Cahn-Hilliard simulation showing tightly packed, small groups of colloids. On the right is a particle simulation with the same number of colloids, showing large loosely packed groupings of colloids.}
\end{figure}

Our model does not accurately capture behavior in the high density regime. In such systems, as ion our 64 and 128 colloid cases, new phenomenon dominate the dynamics. As the entire simulation region becomes enveloped in the reactive regions of the colloids, the solute particles quickly aggregate into either small, high density regions or large regions devoid of colloids bounded by a high density ring. Both of these possibilities are demonstrated in the Figure~\ref{fig:many_coll}. In this regime the colloids are effectively no longer active, as they can no longer localize phase separation around them. Instead we observe large regions of loosely packed colloids. The small clusters of solute particles can be observed to behave like quasi-particles, moving distances similar to those of the individual colloids over a simulation time. In the Cahn-Hilliard model we observe long sinuous bands of tightly packed colloids, with little to no motion of the solute field. Effectively capturing all of these behaviors is outside the capabilities of our simple model. A better starting would likely be the agglomeration model of Smoluchowski~\cite{alma99}.

\section{Conclusions and Future Work}
We  have presented a mesoscale model of an isotropic colloid based on Cahn-Hilliard dynamics. Comparisons of this model against the original particle based model show good qualitative agreement for low numbers of colloids, but quantitatively the enhanced diffusion effect is too large for mid-range densities. Deficiencies in qualitative behavior in our model start to manifest for larger numbers of colloids, where colloids are confined far too quickly compared to the particle model. More work is needed to capture the complicated dynamics of such dense systems. For low to mid-range densities our model has fairly well captured the system behavior at a dramatic speed up in simulation time. Simulating larger or more geometrically complicated colloids is a possible direction for future work with this model, though some care would need to be taken to exclude concentrating solute within the colloid itself in these cases. Non-isotropic mechanisms can also be studied, where $L$ also varies in some non-symmetric way around the colloid.
\section{Acknowledgments and Data Availability}
The authors would like to thank Pierre Illien for helpfully providing the original code used in ~\cite{Deca21}, which served as a useful guide in developing our own code base for this work. The data used in this work and the code used to derive it is available upon request. All code used in this work is available at: https://github.com/david-w-masse/Particle\_Simulation
\section{Appendix A}
\begin{figure}[ht]
\centerline{\includegraphics[scale=0.45,angle=0]{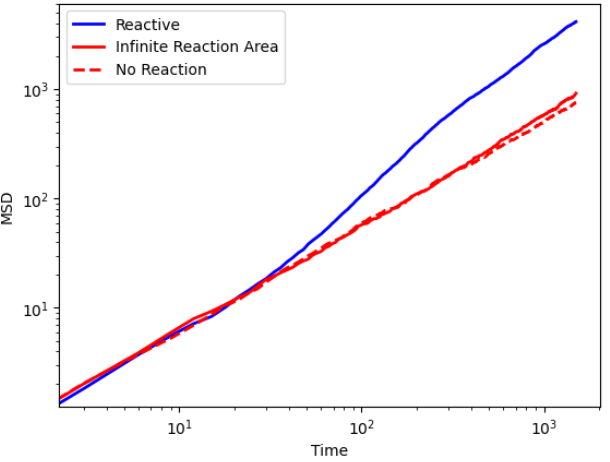}}
\caption{\label{fig:Inactive}\small MSD comparison of an active colloid, an inactive colloid, and a colloid with an infinite reaction radius. This demonstrates that the model parameters must be carefully controlled to achieve the increased diffusion effect.}
\end{figure}
Here we present mean square displacement results for a system with a single inactive colloid, as well as for a single colloid with a reaction radius that is effectively infinite. All other parameters are identical. Compared to a system where the colloid has reaction radius 5, as used in our other simulations, we see the final MSD is less than 1/3 with no enhanced diffusion, and that the characteristic form of an active Brownian particle is lost.


\begin{thebibliography}{10}

\bibitem{Bialke1}
{\sc J.~Bialké, T.~Speck, and H.~Löwen}, {\em Active colloidal suspensions:
  Clustering and phase behavior}, Journal of Non-Crystalline Solids, 407
  (2015), pp.~367--375.
\newblock 7th IDMRCS: Relaxation in Complex Systems.

\bibitem{Bish1}
{\sc K.~J. Bishop, S.~L. Biswal, and B.~Bharti}, {\em Active colloids as
  models, materials, and machines}, Annual Review of Chemical and Biomolecular
  Engineering, 14 (2023), pp.~1--30.

\bibitem{Hill1}
{\sc J.~W. Cahn and J.~E. Hilliard}, {\em {Free Energy of a Nonuniform System.
  I. Interfacial Free Energy}}, The Journal of Chemical Physics, 28 (1958),
  pp.~258--267.

\bibitem{Deca21}
{\sc J.~Decayeux, V.~Dahirel, M.~Jardat, and P.~Illien}, {\em Spontaneous
  propulsion of an isotropic colloid in a phase-separating environment}, Phys.
  Rev. E, 104 (2021), p.~034602.

\bibitem{Deca2}
{\sc J.~Decayeux, M.~Jardat, P.~Illien, and V.~Dahirel}, {\em Conditions for
  the propulsion of a colloid surrounded by a mesoscale phase separation}, The
  European Physical Journal E, 45 (2022), p.~96.

\bibitem{Diaz1}
{\sc J.~D\'{\i}az and I.~Pagonabarraga}, {\em Activity-driven emulsification of
  phase-separating binary mixtures}, Phys. Rev. Lett., 134 (2025), p.~098301.

\bibitem{Duan1}
{\sc Y.~Duan, X.~Zhao, M.~Sun, and H.~Hao}, {\em Research advances in the
  synthesis, application, assembly, and calculation of janus materials},
  Industrial \& Engineering Chemistry Research, 60 (2021), pp.~1071--1095.

\bibitem{Ebens1}
{\sc S.~Ebbens}, {\em Active colloids: Progress and challenges towards
  realising autonomous applications}, Current Opinion in Colloid \& Interface
  Science, 21 (2016), pp.~14--23.

\bibitem{alma99}
{\sc M.~Elimelech}, {\em Particle deposition and aggregation : measurement,
  modelling, and simulation / M. Elimelech ... [et al.].}, Colloid and surface
  engineering series, Butterworth-Heinemann, Oxford [England] ;, 1995.

\bibitem{Ram1}
{\sc R.~Golestanian}, {\em Anomalous diffusion of symmetric and asymmetric
  active colloids}, Phys. Rev. Lett., 102 (2009), p.~188305.

\bibitem{Kaew1}
{\sc C.~Kaewsaneha, P.~Tangboriboonrat, D.~Polpanich, M.~Eissa, and
  A.~Elaissari}, {\em Janus colloidal particles: Preparation, properties, and
  biomedical applications}, ACS Applied Materials \& Interfaces, 5 (2013),
  pp.~1857--1869.
\newblock PMID: 23394306.

\bibitem{Lan1}
{\sc Y.~Lan, M.~Xu, J.~Xie, Y.~Yang, and H.~Jiang}, {\em Spontaneous
  symmetry-breaking of the active cluster drives the directed movement and
  self-sustained oscillation of symmetric rod-like passive particles}, Soft
  Matter, 19 (2023), pp.~3222--3227.

\bibitem{LEE1}
{\sc D.~Lee, J.-Y. Huh, D.~Jeong, J.~Shin, A.~Yun, and J.~Kim}, {\em Physical,
  mathematical, and numerical derivations of the cahn–hilliard equation},
  Computational Materials Science, 81 (2014), pp.~216--225.

\bibitem{Lieb1}
{\sc B.~Liebchen and A.~K. Mukhopadhyay}, {\em Interactions in active
  colloids}, Journal of Physics: Condensed Matter, 34 (2021), p.~083002.

\bibitem{Yama1}
{\sc Y.~research group}, {\em Two-dimensional phase-field model for conserved
  order parameter ({Cahn-Hilliard} equation)}, Self Published,  (2019).

\bibitem{Robert1}
{\sc B.~Robertson, J.~Schofield, P.~Gaspard, and R.~Kapral}, {\em {Molecular
  theory of Langevin dynamics for active self-diffusiophoretic colloids}}, The
  Journal of Chemical Physics, 153 (2020), p.~124104.

\bibitem{Roy1}
{\sc S.~Roy, S.~Dietrich, and A.~Maciolek}, {\em Solvent coarsening around
  colloids driven by temperature gradients}, Phys. Rev. E, 97 (2018),
  p.~042603.

\bibitem{Speck1}
{\sc T.~Speck, J.~Bialk\'e, A.~M. Menzel, and H.~L\"owen}, {\em Effective
  cahn-hilliard equation for the phase separation of active brownian
  particles}, Phys. Rev. Lett., 112 (2014), p.~218304.

\bibitem{Su1}
{\sc H.~Su, C.-A. {Hurd Price}, L.~Jing, Q.~Tian, J.~Liu, and K.~Qian}, {\em
  Janus particles: design, preparation, and biomedical applications}, Materials
  Today Bio, 4 (2019), p.~100033.

\bibitem{Walt1}
{\sc A.~Walther and A.~H.~E. M{\"u}ller}, {\em Janus particles: Synthesis,
  self-assembly, physical properties, and applications}, Chemical Reviews, 113
  (2013), pp.~5194--5261.
\newblock PMID: 23557169.

\bibitem{Zhu1}
{\sc J.~Zhu, L.-Q. Chen, J.~Shen, and V.~Tikare}, {\em Coarsening kinetics from
  a variable-mobility cahn-hilliard equation: Application of a semi-implicit
  fourier spectral method}, Phys. Rev. E, 60 (1999), pp.~3564--3572.

\end{thebibliography}

\end{document}